\documentclass[twocolumn,secnumarabic,amssymb, amsmath, nofootinbib,tightenlines,
nobibnotes, aps, prl,epsfig]{revtex4}
\usepackage{graphicx}
\usepackage{dcolumn}
\usepackage{bm}
\begin{document}
\preprint{APS/123-QED}
\title{Longitudinal structure function from the parton parameterization }

\author{B.Rezaei }
\altaffiliation{brezaei@razi.ac.ir}
\author{G.R.Boroun}%
 \email{grboroun@gmail.com; boroun@razi.ac.ir }
\affiliation{ Physics Department, Razi University, Kermanshah
67149, Iran}
\date{\today}
\begin{abstract}
We present a certain theoretical model to describe data based on
the DGLAP evolution equations at low values of $x$. This model is
based on a hard poemron exchange in the next-to-next-to-leading
order of the perturbative theory. The behavior of the DIS cross
section ratio $R(x,Q^{2})$ and $F_{L}(x,Q^{2})/F_{2}(x,Q^{2})$ is
studied and compared with the experimental data. These behaviors
are controlled by the color dipole model bound. These results show
a good agreement with the DIS experimental data throughout the low
values of $x$. Results can be applied to the LHeC region for
analyses of ultra-high energy processes.\\
\end{abstract}
 \pacs{***}
\keywords{****} 
\maketitle
\subsection{Introduction}

The reduced cross section is defined into the transverse and
longitudinal structure functions, $F_{2}(x,Q^{2})$ and
$F_{L}(x,Q^{2})$, by the following form
\begin{eqnarray}
\widetilde{\sigma}(x,Q^{2})=F_{2}(x,Q^{2})-\frac{y^{2}}{Y_{+}}F_{L}(x,Q^{2}),
\end{eqnarray}
where $Y_{+}=1+(1-y)^2$, $y={Q^{2}}/{xs}$ denotes the inelasticity
and $s$ stands for the center-of-mass  squared energy of  incoming
electrons and protons. As usual $x$ is the Bjorken scaling
parameter and $Q^{2}$ is the four momentum transfer in a deep
inelastic scattering process. For fixed $Q^{2}$, the reduced cross
section rises with decreasing $x$. However, at high-$y$ (very low
$x$) a characteristic bending of the reduced cross section is
observed, which it is attributed to the contribution due to the
longitudinal
structure function.\\
 The longitudinal structure function $F_{L}$ is
proportional to the cross section for the interaction of the
longitudinally polarized virtual photon with a proton. This
observable is interest since it is directly sensitive to the gluon
density. The longitudinal structure function $F_{L}(x,Q^{2})$
obtained as a consequence of the violation of Callan- Gross
relation [1] and is defined as
$F_{L}(x,Q^{2})=F_{2}(x,Q^{2})-2xF_{1}(x,Q^{2})$. Beyond the
parton model the $F_{L}$ effects  can be sizable, hence it can not
be longer neglected. Also, the longitudinal structure function is
predominant in cosmic neutrino-hadron cross section scattering
[2]. This behavior for the longitudinal structure function will be
checked in high energy process such as the Large Hadron electron
Collider (LHeC) project which runs to beyond a TeV in
center-of-mass energy [3].\\
 Data on $F_{L}$ are generally difficult
to extract from reduced cross section measurements. This procedure
require high-precision cross section measurements  at the same
values of $x$ and $Q^{2}$ but at different center-of-mass energy
of the incoming beams [4]. Recently, the new data on the proton
longitudinal structure function has been taken from H1 experiment
at HERA. HERA collected $ep$ collision data with the H1 detector
at a electron beam energy of $27.6~ \mathrm{GeV}$ and  proton beam
energies of $920, 575$ and $460~ \mathrm{GeV}$, which allowed a
measurement of structure functions at $x$ values
$6.5{\times}10^{-4}{\leq}x{\leq}0.65$ and $Q^{2}$
values $35~ \mathrm{GeV}^{2} {\leq}Q^{2}{\leq}800 ~\mathrm{GeV}^{2}$ [5].\\
It is well known that the corresponding Wilson coefficients for
the longitudinal structure functions  were calculated in LO, NLO
and next-to-next-to-leading order (NNLO) [6-8]. Inclusion of
higher order kernels in the study of the longitudinal structure
function is the particular important for studying the physical
processes at new colliders [3]. Authors in Refs.[9-14] presented a
set of formula to extract the longitudinal
  structure function $F_{L}(x,Q^{2})$ from  the transverse structure function $F_{2}$ and its
  derivative  $dF_{2}/d{\ln}Q^{2}$  at low-$x$.\\
  Recently, several methods have been proposed that  the
longitudinal structure function $F_{L}$ can be related to the
nucleon structure function $F_{2}$ and the $F_{2}$ scaling
violation at low values of $x$ [15-17]. These approaches solve the
evolution equations for Laplace and Mellin transform of the parton
densities and subsequently invert the transforms back to
$x$-space, respectively. In Ref.[15], an analytical relation has
been derived for calculating the longitudinal structure function
within the Laplace-transform method at low $x$ values. Moreover,
this method may also use to extract nonliner corrections to the
longitudinal structure function from new collider data. In
Refs.[16-17], authors report  relations between the longitudinal
structure function and the proton structure function based on the
Mellin-transform method at leading order (LO) and next -to-
leading order (NLO) analysis respectively. The obtained results
are based on an analytical solution for the proton structure
function in Refs.[18-19]. They have been suggested a new
parameterization of the proton structure function at asymptotic
low $x$ values. The obtained results indicate a good agreement
with the deep
inelastic scattering (DIS) data on the reduced cross section.\\
In  next section we
 describe the basic theory to extract the longitudinal structure function from the Altarelli- Martinelli
equation [20] and  the distribution functions by solving the DGLAP
evolution equations. Then, the analytical solution of the master
equation  for the longitudinal structure
  function into the gluonic and singlet terms at LO up to NNLO are devoted respectively.
Finally, an analytical analysis of our solutions is presented and
the obtained results are compared with the experimental data which
are
followed by results and discussion.\\
In the present paper  the behavior of the longitudinal structure
function  in $x$-space directly is investigated at LO up to NNLO
analysis. We extend the method using the Regge technique. The
Regge behavior of the singlet and gluon distributions is
introduced by using the power law behavior as
$F_{2}^{s}{\sim}x^{-\lambda_{s}}$ and $G{\sim}x^{-\lambda_{g}}$.
We note that the behavior of distributions with a $Q^{2}$
independent value for exponents obeys the DGLAP equations [21]
when $x^{-\lambda_{s,g}}\gg 1$. This behavior at low-$x$ is well
explained in terms of Regge-like ansatz [22-23]. In this region,
the Regge behavior of the singlet and gluon distribution is
controlled by pomeron exchange. Let us take the power law behavior
for distribution functions as
$F^{s}_{2}(x,Q^{2})=A_{s}(Q^{2})x^{-\lambda_{s}}$ and
$G(x,Q^{2})=A_{g}(Q^{2})x^{-\lambda_{g}}$, that exponents
$\lambda_{s}$ and $\lambda_{g}$ are given as the following
derivatives:
\begin{eqnarray}
\lambda_{s}=\frac{\partial \ln F_{2}^{s}(x,Q^{2})}{\partial
\ln(1/x)},~ \lambda_{g}=\frac{\partial \ln G(x,Q^{2})}{\partial
\ln(1/x)}.
\end{eqnarray}

\subsection{Theory}

The longitudinal structure function  projected from the hadronic
tensor can be considered with combination of the metric and the
spacelike momentum transferred by the virtual photon
$(g_{\mu\nu}-q_{\mu}q_{\nu}/q^{2})$. It is proportional to
hadronic tensor as follows
\begin{equation}
F_{L}(x,Q^{2})/x=\frac{8x^{2}}{Q^{2}}p_{\mu}p_{\nu}W_{\mu\nu}(x,Q^{2}).
\end{equation}
Here $p^{\mu}(p^{\nu})$ is the hadron momentum and $W^{\mu\nu}$ is
the hadronic tensor. The total cross section of a hadronic process
can be written as the sum of the contributions of each parton type
(quarks, antiquarks, and gluons) carrying a fraction of the
hadronic total momentum. In DIS it reads
\begin{equation}
d\sigma_{H}(p)=\sum_{i}{\int}dyd\hat{\sigma}_{i}(yp)\Pi_{i}^{0}(y),
\end{equation}
where $d\hat{\sigma}_{i}$ is the cross section corresponding to
the parton $i$ and $\Pi_{i}^{0}(y)$ is the probability of finding
this parton in the hadron target with the momentum fraction $y$.
Now, taking into account the kinematical constrains one gets the
relation between the hadronic and the partonic structure functions
\begin{eqnarray}
f_{j}(x,Q^{2})&=&\sum_{i}{\int}_{x}^{1}\frac{dy}{y}\textsf{f}_{j}(\frac{x}{y},Q^{2})\Pi_{i}^{0}(y)\\\nonumber
&&=\sum_{i}\textsf{f}_{j}{\otimes}\Pi_{i}^{0}(y)\hspace{0.5cm},j=2,L,
\end{eqnarray}
where $\textsf{f}_{j}(x,Q^{2})=F_{j}(x,Q^{2})/x$ and the symbol
${\otimes}$ denotes a convolution according to the usual
prescription,
$f(x){\otimes}g(x)=\int_{x}^{1}\frac{dy}{y}f(y)g(\frac{x}{y})$.
Equation (5) expresses the hadronic structure functions as the
convolution of the partonic structure function which are
calculable in perturbation theory.\\
In perturbative QCD, the longitudinal structure function in terms
of the coefficient functions is given by [20]
\begin{eqnarray}
x^{-1}F_{L}=C_{L,ns}{\otimes}q_{ns}+<e^{2}>(C_{L,q}{\otimes}q_{s}+C_{L,g}{\otimes}g),
\end{eqnarray}
where the non-singlet quark distribution, $xq_{ns}$, become
negligibly small in comparison with the singlet and gluon
distribution functions, $xq_{s}$ and $xg$, at low values of $x$
and can be ignored. $<e^{k}>$ is the average of the charge $e^{k}$
for the active quark flavors,
$<e^{k}>=n_{f}^{-1}\sum_{i=1}^{n_{f}}e_{i}^{k}$. The perturbative
expansion of the coefficient functions can be written as [7]
\begin{eqnarray}
C_{L,a}(\alpha_{s},x)=\sum_{n=1}a(t)^{n}c_{L,a}^{n}(x),
\end{eqnarray}
where $n$ is the order in the running coupling constant. The
running coupling constant in the high-loop corrections of the
above  equation is expressed entirely  thorough the variable
$a(t)$, as $a(t)=\frac{\alpha_{s}}{4\pi}$. The explicit expression
for the coefficient functions in LO up to NNLO are relegated in
Appendix A.\\
The running coupling constant $\alpha_{s}$ has the following forms
in NLO up to NNLO respectively [24]
\begin{equation}
\alpha_{s}^{\rm
NLO}=\frac{4\pi}{\beta_{0}t}[1-\frac{\beta_{1}{\ln}t}{\beta_{0}^{2}t}],
\end{equation}
and
\begin{eqnarray}
\alpha_{s}^{\rm
NNLO}&=&\frac{4\pi}{\beta_{0}t}[1-\frac{\beta_{1}{\ln}t}{\beta_{0}^{2}t}+\frac{1}{(\beta_{0}t)^{2}}
[(\frac{\beta_{1}}{\beta_{0}})^{2}\nonumber\\
&&(\ln^{2}t-{\ln}t+1)+\frac{\beta_{2}}{\beta_{0}}]].
\end{eqnarray}
where $\beta_{0}=\frac{1}{3}(33-2n_{f})$,
$\beta_{1}=102-\frac{38}{3}n_{f}$ and
$\beta_{2}=\frac{2857}{6}-\frac{6673}{18}n_{f}+\frac{325}{54}n_{f}^{2}$
are one-loop, two-loop and three-loop corrections to the QCD
$\beta$-function and $n_{f}$ denotes the number of effectively
massless flavours. The variable $t$ is defined as
$t={\ln}(\frac{Q^{2}}{\Lambda^{2}})$. We take the $n_{f}=4$ for
$m_{c}^{2}<\mu^{2}<m^{2}_{b}$ and adjust the QCD cut- off
parameter $\Lambda$ at each heavy quark mass threshold,
$\mu^{2}=m^{2}_{c}$. This parameter has been extracted from the
strong coupling constat at the Z boson mass scale for the NLO and
NNLO approximation in accordance with the table III.\\
The Dokshitzer-Gribov-Lipatov-Altarelli-Parisi (DGLAP) [21]
evolution equations are fundamental tools to study the $Q^{2}$-
and $x$-evolutions of structure functions. The structure function
reflects the momentum distributions of the partons in the nucleon.
It is also important to know the gluon distribution inside a
hadron at low-$x$ because gluons are expected to be dominant in
this region. Some analytical solutions of the polarized and
unpolarized DGLAP evolution equations have been reported in last
years [25-26] with considerable phenomenological success. The
flavor-singlet quark density of a hadron is given by
\begin{eqnarray}
q_{s}=\sum_{i=1}^{n_{f}}[q_{i}+\overline{q}_{i}],\nonumber
\end{eqnarray}
where $q_{i}$ and $\overline{q}_{i}$ represent the number
distribution of quarks and antiquarks, respectively, in the
fractional hadron momentum $x$. The subscripts $i$ indicate the
flavor of the quarks and antiquarks. In the common
$\overline{\mathrm{MS}}$ renormalization scheme the proton
structure function, extracted from the DIS ep process, can be
written as the sum of a flavour singlet. Using the fact that the
nonsinglet contribution $F_{2}^{NS}$ can be ignored safely at low
values of $x$ so we will have,
\begin{eqnarray}
\frac{F_{2}(x,Q^{2})}{x}&{\simeq}&\frac{1}{x}(F_{2,s}(x,Q^{2})+F_{2,g}(x,Q^{2}))\nonumber\\
&&=<e^{2}>(C_{2,s}(x,Q^{2}){\otimes}q_{s}(x,Q^{2})\nonumber\\
&&+C_{2,g}(x,Q^{2}){\otimes}g(x,Q^{2})),
\end{eqnarray}
where the corresponding gluon density is denoted by $g(x,Q^{2})$
and $C_{2,i}(i=s,g)$ are the common Wilson coefficient functions.
In the space-like region, the proton structure function can be
expressed in terms of quark and gluon densities
\begin{eqnarray}
F_{2}(x,Q^{2})&=&x\sum_{j=1}^{n_{f}}e_{j}^{2}\int_{x}^{1}\frac{dz}{z}[\frac{1}{n_{f}}q_{s}(\frac{x}{z},\mu_{f}^{2})
C_{2,s}(z,\frac{Q^{2}}{\mu_{f}^{2}})\nonumber\\
&&+g(\frac{x}{z},\mu_{f}^{2})C_{2,g}(z,\frac{Q^{2}}{\mu_{f}^{2}})].
\end{eqnarray}
The scale $\mu_{f}$ denotes the factorization scale. Considering
coupling constant renormalization, the proton structure function
is expressed as follows [27]
\begin{eqnarray}
F_{2}(x,Q^{2})=[C_{2,s}(\alpha_{s}(\mu_{r}^{2}),
\frac{Q^{2}}{\mu_{f}^{2}},\frac{\mu_{f}^{2}}{\mu_{r}^{2}}){\otimes}q_{s}(\alpha_{s}(\mu_{r}^{2}),
\frac{\mu_{f}^{2}}{\mu^{2}},\frac{\mu_{f}^{2}}{\mu_{r}^{2}})\nonumber\\
+C_{2,g}(\alpha_{s}(\mu_{r}^{2}),
\frac{Q^{2}}{\mu_{f}^{2}},\frac{\mu_{f}^{2}}{\mu_{r}^{2}}){\otimes}g(\alpha_{s}(\mu_{r}^{2}),
\frac{\mu_{f}^{2}}{\mu^{2}},\frac{\mu_{f}^{2}}{\mu_{r}^{2}})](x),\nonumber\\
\end{eqnarray}
which the renormalized parton density is defined by the following
form
\begin{eqnarray}
f_{l}(z,\alpha_{s}(\mu_{r}^{2}),
\frac{\mu_{f}^{2}}{\mu^{2}},\frac{\mu_{f}^{2}}{\mu_{r}^{2}})=\sum_{k=q,g}(\Gamma_{lk}(\alpha_{s}(\mu_{r}^{2}),
\frac{\mu_{f}^{2}}{\mu^{2}},\frac{\mu_{f}^{2}}{\mu_{r}^{2}},\epsilon){\otimes}\widehat{f}_{k})(z),\nonumber\\
f_{q}=q_{s}~\mathrm{and}~f_{g}=g~\mathrm{for}~l=q,g.~~~~~~~~~~~~~~~~~~~~~~~~~~~~~~~~~~~~\nonumber
\end{eqnarray}
Here $\mu_{r}$ is the renormalization scale, $\epsilon=n-4$
represents the collinear singularities and $\widehat{f}_{k}$ is
defined as the bare parton density. Due to changes in the scales
$\mu_{r}$ and $\mu_{f}$, the changes in the parton densities and
the coefficient functions can be expressed in terms of the
renormalization group equation [28]. The renormalization group
equation of the parton densities and the scale dependence of the
coefficient functions causes the parton distribution functions to
be expressed in terms of the splitting functions
$P_{ij}(x,\alpha_{s})$. The  evolution equations of distribution
functions in the singlet sector are given by
\begin{widetext}
\begin{eqnarray}
Q^{2}\frac{{\partial}}{{\partial}Q^{2}} \left(\begin{array}{c}
 F^{s}_{2}(x,Q^{2}) \\
G (x,Q^{2})
 \end{array}
 \right)=\left(
 \begin{array}{cc}
P_{qq}(x,\alpha_{s}) & P_{qg}(x,\alpha_{s}) \\
P_{gq}(x,\alpha_{s}) & P_{gg}(x,\alpha_{s})
 \end{array}
 \right){\otimes}
\left(\begin{array}{c}
 F^{s}_{2}(x,Q^{2})\\
G (x,Q^{2})
 \end{array}
 \right),
 \end{eqnarray}
 \end{widetext}
 where $G(x,Q^{2})$ is the gluon distribution function defined
 into the gluon density by $G(x,Q^{2})=xg(x,Q^{2})$.
For the high-order contribution to scaling violation of
distribution functions, we extend the discussion to the NNLO
level. So that the high-order correction to the coupled  DGLAP
evolution equations can be written as
\begin{widetext}
\begin{eqnarray}
Q^{2}\frac{{\partial}}{{\partial}Q^{2}}F_{2}^{s}(x,Q^{2})&=&[\frac{\alpha_{s}(Q^{2})}{4\pi}P_{qq}^{0}(x)
+(\frac{\alpha_{s}(Q^{2})}{4\pi})^{2}P_{qq}^{1}(x)
+(\frac{\alpha_{s}(Q^{2})}{4\pi})^{3}P_{qq}^{2}(x)]{\otimes}F_{2}^{s}(x,Q^{2})\nonumber\\
&&+[\frac{\alpha_{s}(Q^{2})}{4\pi}P_{qg}^{0}(x)+(\frac{\alpha_{s}(Q^{2})}{4\pi})^{2}P_{qg}^{1}(x)
+(\frac{\alpha_{s}(Q^{2})}{4\pi})^{3}P_{qg}^{2}(x)]{\otimes}G(x,Q^{2}).\nonumber\\
Q^{2}\frac{{\partial}}{{\partial}Q^{2}}G(x,Q^{2})&=&[\frac{\alpha_{s}(Q^{2})}{4\pi}P_{gq}^{0}(x)+(\frac{\alpha_{s}(Q^{2})}{4\pi})^{2}P_{gq}^{1}(x)
+(\frac{\alpha_{s}(Q^{2})}{4\pi})^{3}P_{gq}^{2}(x)]{\otimes}F_{2}^{s}(x,Q^{2})\nonumber\\
&&+[\frac{\alpha_{s}(Q^{2})}{4\pi}P_{gg}^{0}(x)
+(\frac{\alpha_{s}(Q^{2})}{4\pi})^{2}P_{gg}^{1}(x)
+(\frac{\alpha_{s}(Q^{2})}{4\pi})^{3}P_{gg}^{2}(x)]{\otimes}G(x,Q^{2}).
\end{eqnarray}
\end{widetext}
The splitting functions $P_{ij}^{n}(n=0,1,2)$  are the Altarelli-
Parisi splitting kernels at one, two and three loops corrections.
The explicit forms of the splitting functions at LO up to NNLO
analysis are given in Appendix B. For brevity the leading-order
contribution to the coupled DGLAP evolution equation can be
written
\begin{widetext}
\begin{eqnarray}
\frac{4\pi}{\alpha_{s}(Q^{2})}\frac{{\partial}G(x,Q^{2})}{{\partial}{\ln}Q^{2}}&=&\frac{33-2n_{f}}{3}G(x,Q^{2})+12G(x,Q^{2}){\ln}\frac{1-x}{x}
+12x{\int_{x}^{1}}\frac{dz}{z-x}(\frac{G({z},Q^{2})}{z}-\frac{G({x},Q^{2})}{x})\nonumber\\
&&+12x{\int_{x}^{1}}G({z},Q^{2})(\frac{z}{x}-2+\frac{x}{z}-\frac{x^{2}}{z^{2}})\frac{dz}{z^{2}}
+\frac{8}{3}{\int_{x}^{1}}F_{2}^{s}({z},Q^{2})(1+(1-\frac{x}{z})^{2})\frac{dz}{z},\nonumber\\
\frac{4\pi}{\alpha_{s}(Q^{2})}\frac{{\partial}F_{2}^{s}(x,Q^{2})}{{\partial}{\ln}Q^{2}}&=&4F_{2}^{s}(x,Q^{2})+\frac{16}{3}F_{2}^{s}(x,Q^{2}){\ln}\frac{1-x}{x}
+\frac{16}{3}x{\int_{x}^{1}}\frac{dz}{z-x}(\frac{F_{2}^{s}({z},Q^{2})}{z}-\frac{F_{2}^{s}({x},Q^{2})}{x})\nonumber\\
&&-\frac{8}{3}x{\int_{x}^{1}}F_{2}^{s}({z},Q^{2})(1+\frac{x}{z})\frac{dz}{z^{2}}
+2n_{f}x{\int_{x}^{1}}G({z},Q^{2})(1-2\frac{x}{z}+2\frac{x^{2}}{z^{2}})\frac{dz}{z}.
\end{eqnarray}
\end{widetext}

Now we consider the behavior of the longitudinal structure
function with respect to the hard-pomeron behavior [22-23,31] by
using the parametrization of $F_{2}(x,Q^{2})$ and $G(x,Q^{2})$
presented in Refs.[19] and [32] respectively. This
study is based on two steps:\\
1) In the first step of the analysis, we only present the
longitudinal structure function behavior by using the gluon
distribution function and its logarithmic
derivative [32].\\
2) Then the longitudinal structure function have been extracted at
low $x$ from the proton structure function $F_{2}(x,Q^{2})$ and
the logarithmic derivative $dF_{2}(x,Q^{2})/d\ln(Q^{2})$ [19].\\
The next-to-leading order corrections are the standard
approximations for the most important processes. The corresponding
one- and two-loop splitting functions have been known for a long
time. As in Refs. [7-8, 29-30,33], the authors have been reported
the complete two- and three-order coefficient functions for the
longitudinal structure functions in deep inelastic scattering
(DIS). The NNLO corrections need to be included, however, in order
to arrive to quantitatively reliable predictions for ultra low
values of $x$ at present and future high-energy colliders. These
corrections are known not only for the distribution functions but
also for the longitudinal coefficient functions in deep-inelastic
scattering (DIS).  Now, a detailed analysis has been performed to
find analytical solutions of the longitudinal structure function
into the distributions and those derivatives,
using the hard-pomeron behavior, at LO up to NNLO approximations.\\

\subsection{Method}

The following parameterization of the deep inelastic scattering
structure function $F_{2}(x,Q^{2})$ defined by
\begin{eqnarray}
F_{2}(x,Q^{2})\sim \sum_{i}A_{i}(Q^{2})x^{-\lambda_{i}}.
\end{eqnarray}
The singlet part of the structure function is controlled by
pomeron exchange. Here the $i=0$ term is hard-pomeron  and $i=1$
is soft-pomeron exchange [21-22]. The effective intercept
behavior, at low values of $x$, exhibiting for the fast growth of
the singlet structure function. The exponent $\lambda_{s}$ is
found to be $\simeq 0.33$ in Refs.[34-35]. It can be recasted into
the symbolic form as
\begin{eqnarray}
F_{2}(x,Q^{2})\propto x^{-\lambda_{s}}.
\end{eqnarray}
The low-$x$ behavior of the gluon distribution function also is
dominated with hard-pomeron intercept as
\begin{eqnarray}
G(x,Q^{2})\propto x^{-\lambda_{g}},
\end{eqnarray}
where   $\lambda_{g}\simeq 0.43$ [22-23,34-35]. Based on the
hard-pomeron behavior for the distribution functions, let us put
Eqs.(17) and (18) in r.h.s of Eqs.(14). After doing the
integration over $z$, These equations  can be rewritten  in a
convolution form as
\begin{eqnarray}
\frac{{\partial}G(x,Q^{2})}{{\partial}{\ln}Q^{2}}&=&G(x,Q^{2})\Phi_{gg}(x,Q^{2})\nonumber\\
&&+F_{2}(x,Q^{2})\Theta_{gq}(x,Q^{2}),
\end{eqnarray}
and
\begin{eqnarray}
\frac{{\partial}F_{2}(x,Q^{2})}{{\partial}{\ln}Q^{2}}&=&F_{2}(x,Q^{2})\Phi_{qq}(x,Q^{2})\nonumber\\
&&+G(x,Q^{2})\Theta_{qg}(x,Q^{2}).
\end{eqnarray}
On the other hand, the Altarelli- Martinelli equation for the
longitudinal stricture function at low $x$ values is given by the
similar method as
\begin{eqnarray}
F_{L}(x,Q^{2})&=&F_{2}(x,Q^{2})I_{L,q}(x,Q^{2})\nonumber\\
&&+G(x,Q^{2})I_{L,g}(x,Q^{2}).
\end{eqnarray}
The analytical results for the compact form of the kernels ($\Phi,
\Theta$ and $I$ ) at LO up to NNLO are given in Appendix C.\\

\subsection{Gluonic Formalism}

Using the above kernels and Eqs.(19) and (21), we can calculate
the $F_{L}$ structure function into the gluon distribution and its
derivative [36-37]. For this purpose the ratio
$\frac{F_{2}(x,Q^{2})}{G(x,Q^{2})}$ is determined by solving the
DGLAP evolution equation for the gluon distribution function
(i.e., Eq.(19)) and the Altarelli- Martinelli equation for the
longitudinal stricture function (i.e. Eq.(21)). Therefore the
$F_{L}$ using the gluonic terms extracted by the following form
\begin{eqnarray}
F_{L}(x,Q^{2})&=&\frac{I_{L,q}(x,Q^{2})}{\Theta_{gq}(x,Q^{2})}\frac{{\partial}G(x,Q^{2})}{{\partial}{\ln}Q^{2}}+\{I_{L,g}(x,Q^{2})\nonumber\\
&&-\Phi_{gg}(x,Q^{2})\frac{I_{L,q}(x,Q^{2})}{\Theta_{gq}(x,Q^{2})}
\}G(x,Q^{2}).
\end{eqnarray}
The accuracy this relation can be checked for certain
parametrization of the gluon distribution function and  its
derivatives.\\
 In the following, we will present our analytical
method based on the newly-proton structure function which strongly
violate the Froissart boundary [38]. We used the hard-pomeron
behavior for the input singlet and gluon exponents to determine
the parton distribution behavior at any $Q^{2}$ values. Having the
exponents and using the proton structure function (i.e., Appendix
D), one can extract the longitudinal structure function as a
function of $x$ at any
desired $Q^{2}$ value.\\

\subsection{Singlet Formalism}

One can rewrite Eqs.(20) and (21) with respect to the proton
structure function $F_{2}(x,Q^{2})$ and its derivative $\partial
F_{2}(x,Q^{2})/\partial{\ln}Q^{2}$. Then we will have
\begin{eqnarray}
F_{L}(x,Q^{2})&=&\frac{I_{L,g}(x,Q^{2})}{\Theta_{qg}(x,Q^{2})}\frac{{\partial}F_{2}(x,Q^{2})}{{\partial}{\ln}Q^{2}}+\{I_{L,q}(x,Q^{2})\nonumber\\
&&-\Phi_{qq}(x,Q^{2})\frac{I_{L,g}(x,Q^{2})}{\Theta_{qg}(x,Q^{2})}
\}F_{2}(x,Q^{2}).
\end{eqnarray}
Eqs.(28-40) (in Appendixes A-C) help to estimate the longitudinal
proton structure function in the leading order up to the
next-to-next-to-leading order approximation, as we get
\begin{widetext}
\begin{eqnarray}
F^{LO}_{L}(x,Q^{2})&=&\frac{1}{2}\frac{c^{LO}_{L,g}(x){\odot}
x^{\lambda_{g}}}{P^{LO}_{qg}(x){\odot} x^{\lambda_{g}}}
\frac{{\partial}F_{2}(x,Q^{2})}{{\partial}{\ln}Q^{2}}+\{\frac{\alpha_{s}}{4\pi}c^{LO}_{L,q}(x){\odot}
x^{\lambda_{s}} -[\frac{\alpha_{s}}{2\pi}P^{LO}_{qq}(x){\odot}
x^{\lambda_{s}}]\frac{1}{2}\frac{c^{LO}_{L,g}(x){\odot}
x^{\lambda_{g}}}{P^{LO}_{qg}(x){\odot} x^{\lambda_{g}}}
\}F_{2}(x,Q^{2}),\nonumber\\
F^{NLO}_{L}(x,Q^{2})&=&\frac{1}{2}\frac{[c^{LO}_{L,g}(x)+\frac{\alpha_{s}}{4\pi}c^{NLO}_{L,g}(x)]{\odot}
x^{\lambda_{g}}}{[P^{LO}_{qg}(x)+\frac{\alpha_{s}}{2\pi}P^{NLO}_{qg}(x)]{\odot}
x^{\lambda_{g}}}
\frac{{\partial}F_{2}(x,Q^{2})}{{\partial}{\ln}Q^{2}}+\{\frac{\alpha_{s}}{4\pi}[c^{LO}_{L,q}(x)+\frac{\alpha_{s}}{4\pi}c^{NLO}_{L,q}(x)]{\odot}
x^{\lambda_{s}}\nonumber\\
&&-[\frac{\alpha_{s}}{2\pi}(P^{LO}_{qq}(x)+\frac{\alpha_{s}}{2\pi}P^{NLO}_{qq}(x)){\odot}
x^{\lambda_{s}}]\frac{1}{2}\frac{[c^{LO}_{L,g}(x)+\frac{\alpha_{s}}{4\pi}c^{NLO}_{L,g}(x)]{\odot}
x^{\lambda_{g}}}{[P^{LO}_{qg}(x)+\frac{\alpha_{s}}{2\pi}P^{NLO}_{qg}(x)]{\odot}
x^{\lambda_{g}}}
\}F_{2}(x,Q^{2}),\nonumber\\
\mathrm{and}\nonumber\\
F^{NNLO}_{L}(x,Q^{2})&=&\frac{1}{2}\frac{[c^{LO}_{L,g}(x)+\frac{\alpha_{s}}{4\pi}c^{NLO}_{L,g}(x)+(\frac{\alpha_{s}}{4\pi})^{2}
c^{NNLO}_{L,g}(x)]{\odot}
x^{\lambda_{g}}}{[P^{LO}_{qg}(x)+\frac{\alpha_{s}}{2\pi}P^{NLO}_{qg}(x)+
(\frac{\alpha_{s}}{2\pi})^{2}P^{NNLO}_{qg}(x) ]{\odot}
x^{\lambda_{g}}}
\frac{{\partial}F_{2}(x,Q^{2})}{{\partial}{\ln}Q^{2}}\nonumber\\
&&+\{\frac{\alpha_{s}}{4\pi}[c^{LO}_{L,q}(x)+\frac{\alpha_{s}}{4\pi}c^{NLO}_{L,q}(x)
+(\frac{\alpha_{s}}{4\pi})^{2}c^{NNLO}_{L,q}(x)]{\odot}
x^{\lambda_{s}}\nonumber\\
&&-[\frac{\alpha_{s}}{2\pi}(P^{LO}_{qq}(x)+\frac{\alpha_{s}}{2\pi}P^{NLO}_{qq}(x)+(\frac{\alpha_{s}}{2\pi})^{2}P^{NNLO}_{qq}(x)){\odot}
x^{\lambda_{s}}]\nonumber\\
&&\frac{1}{2}\frac{[c^{LO}_{L,g}(x)+\frac{\alpha_{s}}{4\pi}c^{NLO}_{L,g}(x)+(\frac{\alpha_{s}}{4\pi})^2c^{NNLO}_{L,g}(x)]{\odot}
x^{\lambda_{g}}}{[P^{LO}_{qg}(x)+\frac{\alpha_{s}}{2\pi}P^{NLO}_{qg}(x)+(\frac{\alpha_{s}}{2\pi})^2P^{NNLO}_{qg}(x)]{\odot}
x^{\lambda_{g}}}
\}F_{2}(x,Q^{2}),\nonumber\\
\end{eqnarray}
\end{widetext}
where we describe the following statement in the convenient form
for further discussion $
f(x){\odot}g(x){\equiv}\int_{x}^{1}(dy/y)f(y)g(y)$.\\
By using the leading-order splitting and coefficient functions
presented in Appendixes A,B and C, the longitudinal structure
function is given by the following form
\begin{widetext}
\begin{eqnarray}
F^{LO}_{L}(x,Q^{2})&=&\frac{12}{5}\frac{\int_{x}^{1}z^{\lambda_{g}+1}dz}{\int_{x}^{1}{(z^2+(1-z)^2)z^{\lambda_{g}}dz}}
\frac{{\partial}F_{2}(x,Q^{2})}{{\partial}{\ln}Q^{2}}[\equiv \mathrm{Derivative~ Eq.(42)~ into~ {\ln}Q^{2}}]\nonumber\\
&&+\{\frac{\alpha_{s}}{4\pi}\int_{x}^{1}8n_{f}(1-z)z^{\lambda_{s}+1}dz
-[\frac{\alpha_{s}}{4\pi}\{4+\frac{16}{3}\ln(\frac{1-x}{x})+\frac{16}{3}\int_{x}^{1}{\frac{z^{\lambda_{s}}-{z}^{-1}}{1-z}dz}\nonumber\\
&&-\frac{8}{3}\int_{x}^{1}{(1+z)z^{\lambda_{s}}dz\}}]\frac{12}{5}\frac{\int_{x}^{1}z^{\lambda_{g}+1}dz}{\int_{x}^{1}{(z^2+(1-z)^2)z^{\lambda_{g}}dz}}
\}F_{2}(x,Q^{2})[\equiv \mathrm{Eq.(42)}].
\end{eqnarray}
\end{widetext}
The NLO and NNLO longitudinal structure functions are too lengthy
to include here and we present them in compact forms of Eqs.(24).
Consequently, one can obtain the longitudinal structure function
from the parameterization of $F_{2}$(i.e., Eq.(42))
at LO, NLO and NNLO approximations respectively.\\

\subsection{Results and Discussion}

In this section, we shall present our results that have been
obtained for the longitudinal structure function $F_{L}(x,Q^{2})$
using the hard-pomeron behavior of the distribution functions to
find an analytical solution for the combined DGLAP and Altarelli-
Martinelli equations. The proton structure function and its
derivative are supposed to be known with respect to the
parameterization represented in Ref.[19]. This parameterization
obtained from a combined fit of the H1 and ZEUS collaborations
data [41] in a range of the kinematical variables $x<0.01$ and
$0.15~\mathrm{GeV^{2}}<Q^{2}<3000~\mathrm{GeV^{2}}$.\\
We use the values of $\lambda_{s}\simeq 0.33$ and
$\lambda_{g}\simeq 0.43$ within the range of $Q^{2}$ under study
[34-35, 42-43]. The coupling constant defined  via the $n_{f}=4$
definition of $\Lambda_{QCD}$ for the ZEUS data [41] and the MRST
set of partons [44]. The values of $\Lambda_{QCD}$ at LO up to
NNLO
 are displayed in Table III respectively. The predictions for  the longitudinal structure
function, in the kinematic range where it has been measured by
$H1$
collaboration [5], are computed and compared at low values of $x$.\\
The results at LO up to NNLO are presented in Fig.1 and compared
with the $H1$ data [5]. In comparison with  Refs.[16-17], the
results have been depicted at fixed value of the invariant mass W
(i.e. $W=230~ \mathrm{GeV}$). As can be seen in this figure, the
results are comparable with the H1 data in  the interval
$1~\mathrm{GeV^{2}}<Q^{2}<500~\mathrm{GeV^{2}}$ in both NLO and
NNLO analysis. At all $Q^{2}$ values the NNLO extracted
longitudinal structure functions are in a good agreement with
experimental data. The $x$-evolution results of $F_{L}$ structure
function are depicted in Fig.2 where we have compared these
results at LO, NLO and NNLO with H1 data. As can be seen in all
figures, the increase of these calculations for the longitudinal
structure function $F_{L}(x,Q^{2})$ towards low $x$ are consistent
with theoretical investigations of ultra-high energy processes
with cosmic neutrinos. Also the NNLO results are in the context of
the
Froissart bound at very low $x$ values.\\
In what follows the ratio $F_{L}(x,Q^{2})/F_{2}(x,Q^{2})$ is
calculated and presented in Figs.3 and 4. Here we use directly the
parametrization of $F_{2}$ from Ref.[19]  in the ratio
$\frac{F_{L}}{F_{2}}$. In these figures the ratio of the structure
functions are compared with the H1 data [5]. The error bars of the
ratio $\frac{F_{L}}{F_{2}}$ are determined by the following form
[45]
\begin{eqnarray}
\Delta(\frac{F_{L}}{F_{2}})=\frac{F_{L}}{F_{2}}\sqrt{(\frac{\Delta
F_{L} }{F_{L}})^{2} +(\frac{\Delta F_{2} }{F_{2}})^{2}},
\end{eqnarray}
where $\Delta F_{L}$ and  $\Delta F_{2}$ are collected from the H1
experimental data in Ref.[5]. The NNLO result obtained of the
ratio $\frac{F_{L}}{F_{2}}$ is comparable with the color dipole
model bound [46].
 The good agreement  between this
method at NNLO analysis and the experimental data indicates that
these results have a bound asymptotic behavior and it is
compatible with the color dipole model bound (
$\frac{F_{L}(x,Q^{2})}{F_{2}(x,Q^{2})}< 0.27$). In Fig.4 we
observe that the ratio at NNLO is almost independent of
$x$-evolution. This strict bound is comparable with the color
dipole bound and experimental data in $x-Q^{2}$ domain.\\
The measurements of longitudinal structure function $F_{L}$ are
used to determine the DIS cross section ratio $R(x,Q^{2})$. The
value of $R$ depends on the ratio of the longitudinal to
transversal cross sections, as
\begin{eqnarray}
R(x,Q^{2})&=&\frac{\sigma_{L}(x,Q^{2})}{\sigma_{T}(x,Q^{2})}\nonumber\\
&&=\frac{F_{L}(x,Q^{2})}{F_{2}(x,Q^{2})-F_{L}(x,Q^{2})},
\end{eqnarray}
which $\sigma_{L}$ and $\sigma_{T}$ are the absorption cross
section of longitudinally and transversely polarized virtual
photon by proton. This ratio is expected to vanish at large
$Q^{2}$ and moderate $x$ in the naive parton model, but is nonzero
at low values of $x$. It dues to the fact that partons can carry
transverse momentum [47]. In Figs.(5) and (6) we have also studied
the behavior of the ratio $R$ at low $x$ in LO, NLO and NNLO using
the calculated values of $F_{L}$ and known parametrized
$F_{2}$ structure function.\\
In Fig.(5) we present the ratio $R$ in comparison with the
 H1 data at LO up to NNLO approximations. As can be seen in this
 figure, one can conclude that the NNLO results essentially improve the good agreement with data
 in comparison with the NLO calculations. The ratio $R$ value
 determined at NNLO is comparable with those obtained in
 literature (of the order of 0.1 to 0.3).\\
  As in Ref.[48] the ratio $R$ is found at
$R=0.260{\pm}0.050$ which this value is constant at the region
$7.10^{-5}<x<2.10^{-3}$ and $3.5\leq Q^{2} \leq 45~
\mathrm{GeV^{2}}$.  In color dipole model the ratio $R$ lead to
the bound $R\leq 0.372$ [49]. In Ref.[50] ZEUS Collaboration is
shown that the overall value of $R$ from both the unconstrained
and constrained fits is $R = 0.105^{+0.055} _{-0.037}$ in wide
range of $Q^{2}$ values ($5\leq Q^{2} \leq
110~GeV^{2}$).\\
In Fig.(6), the ratio $R$ is plotted against $x$ for different
values of $Q^{2}$ in comparison with the H1 data. We have analyzed
the behavior of the ratio $R$ for $Q^{2}$ values of
$Q^{2}=5,~15,~25$ and $45~\mathrm{GeV^{2}}$ in the range
$10^{-5}<x<10^{-2}$ which show a good agreement with the H1 data.
We observe that the ratio $R$ is consistent with a constant
behavior with respect to $x$ for fixed values of $Q^{2}$ at NNLO approximation.\\
To emphasize the size of the NNLO corrections, we show  the ratio
NNLO/NLO for the longitudinal structure function $F_{L}$ and the
ratio $R$ in Figs.(7) and (8). As can be seen, these corrections
are determined in the interval
$1~\mathrm{GeV^{2}}<Q^{2}<5000~\mathrm{GeV^{2}}$ and for
$Q^{2}=5,15,25$ and $45~\mathrm{GeV^{2}}$ respectively. In Fig.(7)
the results for higher values of $Q^{2}$
($Q^{2}>300~\mathrm{GeV^{2}}$) are very similar. It is seen that
the ratio NNLO/NLO for the longitudinal structure function $F_{L}$
and the ratio $R$ has not a continuous increase towards large
$Q^{2}$ values. In this region a depletion is observable, as the
NNLO distributions are smaller in contrast with the NLO
distributions. A detailed comparison for the size of the NNLO
corrections at four $Q^{2}$ values has  been shown in Fig.8. It is
observable that
the corrections at very low values of $x$ are larger than to the moderate $x$ values.\\
\subsection{Summary and Conclusion}

We presented a certain theoretical model to describe the
longitudinal structure function data based on the DGLAP evolution
equations at low values of $x$ in the leading-order up to
next-to-next-to-leading order. The longitudinal structure
functions are arising from the coupled DGLAP and Altarelli-
Martinelli equations by using the hard-pomeron behavior of the
distribution functions. The direct extraction of $F_{L}$ from
experimental data is a cumbersome procedure, therefore the
possibility of the non-direct determination of $F_{L}$ provided by
these equations. This method can be used in very low $x$ at the
LHeC project. Our results at NNLO are in good agreement with the
experimental data at low $x$ values in a large interval of the
momentum transfer. The obtained explicit expression for the
longitudinal structure function is entirely determined by the
effective exponents of the singlet and gluon distribution
functions. We have also calculated the ratio $F_{L}/F_{2}$ and the
cross section ratio $R$ which indicate a good agreement with the
H1 data at the NNLO approximation. The variations of the ratio
$F_{L}/F_{2}$ and $R$ with low values of $x$ and fixed $Q^{2}$
value show a constant behavior similar to that of the experimental
data and the color dipole model. Consequently, the size of the
NNLO corrections become possible to perform the high-order
corrections to the ultra-high energy
processes.\\

\subsection{ACKNOWLEDGMENTS}

The authors are thankful to the Razi University for financial
support of this project.\\

\subsection{Appendix A}

The coefficient functions for $F_{L}$ have the following forms
[7] as:\\
At LO :\\
\begin{eqnarray}
c^{1}_{L,g}&=&8n_{f}z(1-z),\nonumber\\
c^{1}_{L,q}&=&4C_{F}z.
\end{eqnarray}
At NLO :\\
\begin{eqnarray}
c^{2}_{L,g}&=&n_{f}\{(94.74-49.20z)z_{1}L_{1}^2+864.8z_{1}L_{1}\nonumber\\
&&+1161zL_{1}L_{0}+60.06zL_{0}^2+39.66z_{1}L_{0}\nonumber\\
&&-5.333(1/z-1)\},\nonumber\\
c^{2}_{L,q}&=&128/9zL_{1}^2-46.50zL_{1}-84.094L_{0}L_{1}-37.338\nonumber\\
&&+89.53z+33.82z^2+zL_{0}(32.90+18.41L_{0})\nonumber\\
&&-128/9L_{0}-0.012\delta(z_{1})+16/27n_{f}(6zL_{1}\nonumber\\
&&-12zL_{0}-25z+6)\nonumber\\
&&+n_{f}\{(15.94-5.212z)z_{1}^2L_{1}+(0.421+1.520z)L_{0}^2\nonumber\\
&&+28.09z_{1}L_{0}-(2.371/z-19.27)z_{1}^3\}.
\end{eqnarray}
At NNLO :\\
\begin{eqnarray}
c^{3}_{L,g}&=&n_{f}\{(144L_{1}^4-47024/27L_{1}^3+6319L_{1}^2+53160L_{1})z_{1}\nonumber\\
&&+72549L_{0}L_{1}+88238L_{0}^2L_{1}+(3709-33514z\nonumber\\
&&-9533z^2)z_{1}+66773zL_{0}^2-1117L_{0}+45.37L_{0}^2\nonumber\\
&&-5360/27L_{0}^3-(2044.70z_{1}+409.506L_{0})1/z\}\nonumber\\
&&+n_{f}^2\{(32/3L_{1}^3-1216/9L_{1}^2-592.3L_{1}\nonumber\\
&&+1511zL_{1})z_{1}+311.3L_{0}L_{1}+14.24L_{0}^2L_{1}\nonumber\\
&&+(577.3-729z)z_{1}+30.78zL_{0}^3+366L_{0}+1000/9L_{0}^2\nonumber\\
&&+160/9L_{0}^3+88.50371/zz_{1}\}\nonumber\\
&&+fl_{11}^{g}n_{f}^2\{(-0.0105L_{1}^3+1.550L_{1}^2+19.72zL_{1}\nonumber\\
&&-66.745z+0.615z^2)z_{1}+20/27zL_{0}^4+(280/81\nonumber\\
&&+2.260z)zL_{0}^3-(15.40-2.201z)zL_{0}^2+2.260z)zL_{0}^3\nonumber\\
&&-(71.66-0.121z)zL_{0}\},\nonumber\\
c^{3}_{L,q}&=&512/27L_{1}^4-177.40L_{1}^3+650.6L_{1}^2-2729L_{1}\nonumber\\
&&-2220.5-7884z+4168z^2-(844.7L_{0}\nonumber\\
&&+517.3L_{1})L_{0}L_{1}+(195.6L_{1}-125.3)z_{1}L_{1}^3\nonumber\\
&&+208.3zL_{0}^3-1355.7L_{0}-7456/27L_{0}^2
-1280/81L_{0}^3\nonumber\\
&&+0.113\delta(z_{1})+n_{f}\{1024/81L_{1}^3-112.35L_{1}^2+344.1L_{1}\nonumber\\
&&+408.4-9.345z
-919.3z^2+(239.7+20.63L_{1})z_{1}L_{1}^2\nonumber\\
&&+(887.3+294.5L_{0}-59.14L_{1})L_{0}L_{1}-1792/81zL_{0}^3\nonumber\\
&&+200.73L_{0}+64/3L_{0}^2+0.006\delta(z_{1})\}+n_{f}^2\{3zL_{1}^2\nonumber\\
&&+(6-25z)L_{1}-19+(317/6-12\zeta_{2})z
-6zL_{0}L_{1}\nonumber\\
&&+6zLi_{2}(x))+9zL_{0}^2-(6-50z)L_{0}\}64/81\nonumber\\
&&+fl_{11}^{ns}n_{f}\{(107+321.05z-54.62z^2)z_{1}-26.717\nonumber\\
&&+9.773L_{0}+(363.8+68.32L_{0})zL_{0}\nonumber\\
&&-320/81L_{0}^2(2+L_{0})\}z\nonumber\\
&&+n_{f}\{(1568/27L_{1}^3-3968/9L_{1}^2+5124L_{1})z_{1}^2\nonumber\\
&&+(2184L_{0}+6059z_{1})L_{0}L_{1}
-(795.6+1036z)z_{1}^2\nonumber\\
&&-143.6z_{1}L_{0}+2848/9L_{0}^2-1600/27L_{0}^3\nonumber\\
&&-(885.53z_{1}+182L_{0})1/zz_{1}\}
+n_{f}^2\{(-32/9L_{1}^2\nonumber\\
&&+29.52L_{1})z_{1}^2+(35.18L_{0}+73.06z_{1})L_{0}L_{1}\nonumber\\
&&-35.24zL_{0}^2-(14.16-69.84z)z_{1}^2
-69.41z_{1}L_{0}\nonumber\\
&&-128/9L_{0}^2+40.2391/zz_{1}^2\}+fl_{11}^{ps}n_{f}\{(107\nonumber\\
&&+321.05z-54.62z^2)z_{1}-26.717+9.773L_{0}+(363.8\nonumber\\
&&+68.32L_{0})zL_{0}-320/81L_{0}^2(2+L_{0})\}z.
\end{eqnarray}
In these equations we have used the abbreviations
$z_{1}=1-z,~L_{0}=\ln{z}$ and $L_{1}=\ln{z_{1}}$. For the SU(N)
gauge group, we have $C_{A}=N$, $C_{F}=(N^{2}-1)/2N$,
 $T_{F}=n_{f}T_{R}$, and $T_{R}=1/2$ where $C_{F}$ and $C_{A}$ are the color Cassimir
 operators in QCD. Also the new charge factors are defined by the following
form, $fl_{11}^{ns}=3<e>$, $fl_{11}^{g}=<e>^{2}/<e^{2}>$ and
$fl_{11}^{ps}=fl_{11}^{g}-fl_{11}^{ns}$.\\

\subsection{Appendix B}

The LO up to NNLO splitting functions for singlet and gluon
distribution functions are as follows [29-30]:
At LO :\\
\begin{eqnarray}
&&P^{\rm LO}_{qq}(z)=C_{F}[\frac{1+z^{2}}{(1-z)_{+}}+\frac{3}{2}\delta(1-z)].\nonumber\\
&&P^{\rm LO}_{qg}(z)=\frac{1}{2}(z^{2}+(1-z)^{2}).\nonumber\\
&&P^{\rm LO}_{gq}(z)=C_{F}\frac{1+(1-z)^2}{z}.\nonumber\\
&&P^{\rm LO}_{gg}(z)=2C_{A}(\frac{z}{(1-z)_{+}}+\frac{(1-z)}{z}+z(1-z))\nonumber\\
&&+\delta(1-z)\frac{(11C_{A}-4n_{f}T_{R})}{6}.
\end{eqnarray}
 The convolution integrals  which
contains plus prescription, $()_{+}$, can be easily calculate by
\begin{eqnarray}
\int_{x}^{1}\frac{dy}{y}f(\frac{x}{y})_{+}g(y)&=&\int_{x}^{1}\frac{dy}{y}f(\frac{x}{y})[g(y)-\frac{x}{y}g(x)]\nonumber\\
&&-g(x)\int_{0}^{x}f(y)dy\nonumber\\
\end{eqnarray}
At NLO :\\
\begin{widetext}
\begin{eqnarray}
P_{qq}^{\rm NLO}&=&(C_F)^2(-1+z+(1/2-3/2z)\ln(z)-1/2(1+z)\ln(z)^{2}-(3/2\ln(z)+2\ln(z)\ln(1-z))p_{qq}(z)\nonumber\\
&&+2p_{qq}(-z)S_2(z))+C_F C_A(14/3(1-z)+(11/6\ln(z)+1/2\ln(z)^{2}+67/18-\pi^2/6)p_{qq}(z)\nonumber\\
&&-p_{qq}(-z)S_2(z))+C_FT_F(-16/3+40/3z+(10z+16/3z^2+2)\ln(z)-112/9z^2+40/(9z)\nonumber\\
&&-2(1+z)\ln(z)^{2}-(10/9+2/3\ln(z))p_{qq}(z)).\nonumber\\
P_{qg}^{\rm NLO}&=&C_FT_F(4-9z-(1-4z)\ln(z)-(1-2z)\ln(z)^{2}+4\ln(1-z)+(2\ln((1-z)/z)^{2}-4ln((1-z)/z)\nonumber\\
&&-2/3\pi^2+10)P_{qg}(z))+C_AT_F(182/9+14/9z+40/(9z)+(136/3z-38/3)\ln(z)-4\ln(1-z)\nonumber\\
&&-(2+8z)\ln(z)^{2}+2P_{qg}(-z)S_2(z)+(-\ln(z)^{2}+44/3\ln(z)-2\ln(1-z)^{2}+4\ln(1-z)\nonumber\\
&&+\pi^2/3-218/9)P_{qg}(z)).\nonumber\\
P_{gq}^{\rm NLO}&=&C_F^2(-5/2-7z/2+(2+7/2z)\ln(z)-(1-z/2)ln(z)^{2}-2z\ln(1-z)-(3\ln(1-z)\nonumber\\
&&+\ln(1-z)^{2})P_{gq}(z))+C_FC_A(28/9+65/18z+44/9z^2-(12+5z+8/3z^2)\ln(z)+(4+z)\ln(z)^{2}\nonumber\\
&&+2z\ln(1-z)+S_2(z)P_{gq}(-z)+(1/2-2\ln(z)\ln(1-z)+1/2\ln(z)^{2}+11/3\ln(1-z)+\ln(1-z)^{2}\nonumber\\
&&-\pi^2/6)P_{gq}(z))+C_FT_F(-4/3z-(20/9+4/3\ln(1-z))P_{gq}(z)).\nonumber\\
P_{gg}^{\rm NLO}&=&C_FT_F(-16+8z+20/3z^2+4/(3z)-(6+10z)\ln(z)-(2+2z)\ln(z)^{2})+C_AT_F(2-2z\nonumber\\
&&+26/9(z^2-1/z)-4/3(1+z)\ln(z)-20/9P_{gg}(z))+C_A^2(27/2(1-z)+26/9(z^2-1/z)\nonumber\\
&&-(25/3-11/3z+44/3z^2)\ln(z)+4(1+z)\ln(z)^{2}+2P_{gg}(-z)S_2(z)+(67/9-4\ln(z)\ln(1-z)\nonumber\\
&&+\ln(z)^{2}-\pi^2/3)P_{gg}(z)).\nonumber\\
\end{eqnarray}
\end{widetext}
where
\begin{eqnarray}
p_{qq}(z)=2/(1-z)-1-z\nonumber\\
p_{qq}(-z)=2/(1+z)-1+z\nonumber\\
P_{qg}(z)=z^2+(1-z)^2\nonumber\\
P_{qg}(-z)=z^2+(1+z)^2\nonumber\\
P_{gq}(z)=(1+(1-z)^2)/z\nonumber\\
P_{gq}(-z)=-(1+(1+z)^2)/z\nonumber\\
P_{gg}(z)=1/(1-z)+1/z-2+z(1-z)\nonumber\\
P_{gg}(-z)=1/(1+z)-1/z-2-z(1+z)\nonumber\\
S_2(z)=\int_{1/(1+z)}^{z/(1+z)}1/y\ln((1-y)/y)dy\nonumber\\
\end{eqnarray}
At NNLO :\\
\begin{widetext}
\begin{eqnarray}
P_{qq}^{\rm NNLO}&=&(n_f(-5.926L_{1}^3-9.751L_{1}^2-72.11L_{1}+177.4+392.9z-101.4z^2-57.04L_{0}L_{1}-661.6L_{0}\nonumber\\
&&+131.4L_{0}^2-400/9L_{0}^3+160/27L_{0}^4-506/z-3584/271/zL_{0})+n_f^2(1.778L_{1}^2+5.944L_{1}\nonumber\\
&&+100.1-125.2z+49.26z^2-12.59z^3-1.889L_{0}L_{1}+61.75L_{0}+17.89L_{0}^2+32/27L_{0}^3\nonumber\\
&&+256/811/z))(1-z).\nonumber\\
P_{qg}^{\rm NNLO}&=&n_f(100/27L_{1}^4-70/9L_{1}^3-120.5L_{1}^2+104.42L_{1}+2522-3316z+2126z^2\nonumber\\
&&+L_{0}L_{1}(1823-25.22L_{0})-252.5zL_{0}^3+424.9L_{0}+881.5L_{0}^2-44/3L_{0}^3+536/27L_{0}^4\nonumber\\
&&-1268.31/z-896/31/zL_{0})+n_f^2(20/27L_{1}^3+200/27L_{1}^2-5.496L_{1}-252+158z+145.4z^2\nonumber\\
&&-139.28z^3-L_{0}L_{1}(53.09+80.616L_{0})-98.07zL_{0}^2+11.70zL_{0}^3-254L_{0}-98.80L_{0}^2-376/27L_{0}^3\nonumber\\
&&-16/9L_{0}^4+1112/2431/z).\nonumber\\
P_{gq}^{\rm NNLO}&=&400/81L_{1}^4+2200/27L_{1}^3+606.3L_{1}^2+2193L_{1}-4307+489.3z+1452z^2+146z^3-447.3L_{0}^2L_{1}\nonumber\\
&&-972.9zL_{0}^2+4033L_{0}-1794L_{0}^2+1568/9L_{0}^3-4288/81L_{0}^4+6163.11/z+1189.31/zL_{0}\nonumber\\
&&+n_f(-400/81L_{1}^3-68.069L_{1}^2-296.7L_{1}-183.8+33.35z-277.9z^2+108.6zL_{0}^2\nonumber\\
&&-49.68L_{0}L_{1}+174.8L_{0}+20.39L_{0}^2+704/81L_{0}^3+128/27L_{0}^4-46.411/z+71.0821/zL_{0})\nonumber\\
&&+n_f^2(96/27L_{1}^2(1/z-1+1/2z)+320/27L_{1}(1/z-1+4/5z)-64/27(1/z-1-2z)).\nonumber\\
P_{gg}^{\rm NNLO}&=&2643.524D_{0}+4425.894\delta(1-z)+3589L_{1}-20852+3968z-3363z^2+4848z^3\nonumber\\
&&+L_{0}L_{1}(7305+8757L_{0})+274.4L_{0}-7471L_{0}^2+72L_{0}^3-144L_{0}^4+142141/z+2675.81/zL_{0}\nonumber\\
&&+n_f(-412.142D_{0}-528.723\delta(1-z)-320L_{1}-350.2+755.7z-713.8z^2+559.3z^3\nonumber\\
&&+L_{0}L_{1}(26.15-808.7L_{0})+1541L_{0}+491.3L_{0}^2+832/9L_{0}^3+512/27L_{0}^4+182.961/z\nonumber\\
&&+157.271/zL_{0})+n_f^2(-16/9D_{0}+6.4630\delta(1-z)-13.878+153.4z-187.7z^2+52.75z^3\nonumber\\
&&-L_{0}L_{1}(115.6-85.25z+63.23L_{0})-3.422L_{0}+9.680L_{0}^2-32/27L_{0}^3-680/2431/z).
\end{eqnarray}
\end{widetext}
where $D_{0}=1/(1-z)$.\\

\subsection{Appendix C}

We present here the kernels for the quark and gluon sectors,
denoted by $\Phi$ and $\Theta$ respectively at LO up to NNLO which
we used in Eqs.(19-21).\\
\begin{eqnarray}
\Theta_{gq}(x,Q^{2})&=&P_{gq}(x,\alpha_{s}){\odot}
x^{\lambda_{s}},\nonumber\\
\Theta_{qg}(x,Q^{2})&=&P_{qg}(x,\alpha_{s}){\odot}
x^{\lambda_{g}},\nonumber\\
\Phi_{gg}(x,Q^{2})&=&P_{gg}(x,\alpha_{s}){\odot}
x^{\lambda_{g}},\nonumber\\
\Phi_{qq}(x,Q^{2})&=&P_{qq}(x,\alpha_{s}){\odot} x^{\lambda_{s}},
\end{eqnarray}
where the splitting functions expanded into one, two and three
loops correction in accordance with  Appendix B. As the required
leading order approximation of the singlet kernels are written as:
\begin{eqnarray}
\Phi_{qq}(x,Q^{2})&=&\frac{\alpha_{s}}{4\pi}\{4+\frac{16}{3}\ln(\frac{1-x}{x})+\frac{16}{3}\int_{x}^{1}{\frac{z^{\lambda_{s}}-{z}^{-1}}{1-z}dz}\nonumber\\
&&-\frac{8}{3}\int_{x}^{1}{(1+z)z^{\lambda_{s}}dz\}},\nonumber\\
\Theta_{qg}(x,Q^{2})&=&\frac{\alpha_{s}}{4\pi}\frac{20}{9}\int_{x}^{1}{(z^2+(1-z)^2)z^{\lambda_{g}}dz}.
\end{eqnarray}
The longitudinal kernels at low-$x$ limit are given by
\begin{eqnarray}
I_{L,q}(x,Q^{2})&=&C_{L,q}(x,\alpha_{s}){\odot}
x^{\lambda_{s}},\nonumber\\
I_{L,g}(x,Q^{2})&=&C_{L,g}(x,\alpha_{s}){\odot} x^{\lambda_{g}}.
\end{eqnarray}
Here the longitudinal coefficient functions
$C_{L,a}(x,\alpha_{s}),~a=q,~g$ are the singlet and gluon
functions known perturbatively upto first few orders in the
running coupling constant $\alpha_{s}$ and can be written as
\begin{eqnarray}
C_{L,a}(x,\alpha_{s})&=&\frac{\alpha_{s}}{4\pi}c^{1}_{L,a}(x)+(\frac{\alpha_{s}}{4\pi})^{2} c^{2}_{L,a}(x)\nonumber\\
&&+(\frac{\alpha_{s}}{4\pi})^{3} c^{3}_{L,a}(x),
\end{eqnarray}
where $c_{L,a}(x)$ are given in appendix A. The required leading
order approximation of the longitudinal singlet and gluon
coefficient functions are written as:
\begin{eqnarray}
I_{L,q}(x,Q^{2})&=&\frac{\alpha_{s}}{4\pi}\int_{x}^{1}8n_{f}(1-z)z^{\lambda_{s}+1}dz,\nonumber\\
I_{L,g}(x,Q^{2})&=&\frac{\alpha_{s}}{4\pi}\int_{x}^{1}4C_{F}z^{\lambda_{g}+1}dz.
\end{eqnarray}
\subsection{Appendix D}
The proton structure function parameterized with a  global fit
function [39] to the HERA combined data for $F^{\gamma p}_{ 2}
(x,Q^{2})$ for $0.85< Q^{2} < 3000~ GeV^{2}$ and $x< 0.1$, which
ensures that the saturated Froissart $\ln^{2}(1/x)$ behavior
dominates at low-$x$. This global fit takes the form
\begin{eqnarray}
F^{\gamma p}_{ 2} (x,Q^{2})& =& (1- x)[\frac{F_{P}}{1 -x_{P}} +
A(Q^{2})\ln( \frac{x_{P}}{ x}\frac{1 - x}{ 1 - x_{P}})\nonumber\\
&&+B(Q^{2}) \ln^{2}( \frac{x_{P}}{ x}\frac{1 - x}{ 1 - x_{P}})],
\end{eqnarray}
where
\begin{eqnarray}
 A(Q^{2}) = a_{0} + a_{1} {\ln}Q^{2} + a_{2} {\ln}^{2}
 Q^{2},\nonumber
 \end{eqnarray}
and
\begin{eqnarray}
  B(Q^{2}) = b_{0} + b_{1}
{\ln}Q^{2} + b_{2} {\ln}^{2} Q^{2}.\nonumber
\end{eqnarray}
The fitted parameters are tabulated in Table I. At low $x$ ( or
large $\nu = \ln(1/x)$), the global fit becomes a quadratic
polynomial in $\nu$ as

$ \widehat{F}^{\gamma p}_{ 2} (\nu,Q^{2}){\rightarrow} C_{0f}
(Q^{2}) +  C_{1f} (Q^{2})\nu + C_{2f} (Q^{2})\nu^{2} +
\widehat{O}(\nu)$ where the coefficient functions are defined in
Ref. [39].\\
\begin{table}[h]
\caption{ Parameters of Eq. (41), resulting from a global fit to
the HERA combined data.}
\begin{tabular} {cccc}
\toprule \\  \multicolumn{2}{c}{parameters \quad \quad \quad ~~~~~~~~~~~~~~~~value}    \\ &&&\\ \hline \\ &&&\\
  $a_{0} $  &   \quad  $-8.471\times 10^{-2}\pm2.62\times10^{-3} $  \\
  $a_{1} $  &   \quad   $4.190\times 10^{-2}\quad\pm1.56\times10^{-3}$  \\
  $a_{2}$   &    \quad  $-3.976\times 10^{-3}\pm2.13\times10^{-4}$   \\  &&&\\
 $b_{0}$   &   \quad   $1.292\times 10^{-2}\pm3.62\times10^{-4} $ \\

 $b_{1}$   &   \quad   $2.473\times 10^{-4}\pm2.46\times10^{-4}$  \\

 $b_{2}$    &    \quad  $1.642\times 10^{-3}\pm5.52\times10^{-5} $ \\ &&& \\
$F_{p}$& \quad  $0.413\pm 0.003$ & &\\

$\chi^{2}(\mathrm{goodness~ of~ fit})$ &  \quad  $1.17$ & &\\
\hline

\end{tabular}
\end{table}

This form for $F_{2}^{{\gamma}p}$ (i.e.Eq.(41)) describes the HERA
data well, but the model does not have the properties necessary
for the $\gamma^{*}-p$ reduced cross section to extend smoothly to
$Q^{2}=0$ limit. Authors in Ref.[19] provide good fits to the HERA
data at low $x$ and large $Q^{2}$ values. With respect to the
Block-Halzen  [40] fit, The explicit expression for the proton
structure function in a range of the kinematical variables $x$ and
$Q^{2}$, $x<0.001$ and
$0.15~\mathrm{GeV}^{2}<Q^{2}<3000~\mathrm{GeV}^{2}$, is defined by
the following form
\begin{eqnarray}
F^{\gamma p}_{ 2}(x,Q^{2})& =& D(Q^{2})(1-
x)^{n}[C(Q^{2})+A(Q^{2})\ln(\frac{1}{x}\frac{Q^{2}}{Q^{2}+\mu^{2}})\nonumber\\
&&+B(Q^{2})\ln^{2}(\frac{1}{x}\frac{Q^{2}}{Q^{2}+\mu^{2}})],
\end{eqnarray}
where
\begin{eqnarray}
 A(Q^{2})& =& a_{0} + a_{1} {\ln}(1+\frac{Q^{2}}{\mu^{2}}) + a_{2}{\ln}^{2}(1+\frac{Q^{2}}{\mu^{2}})
 ,\nonumber\\
B(Q^{2})& =& b_{0} + b_{1} {\ln}(1+\frac{Q^{2}}{\mu^{2}}) +
b_{2}{\ln}^{2}(1+\frac{Q^{2}}{\mu^{2}})
 ,\nonumber\\
C(Q^{2})& =& c_{0} + c_{1}
{\ln}(1+\frac{Q^{2}}{\mu^{2}}),\nonumber\\
D(Q^{2})& =& \frac{Q^{2}(Q^{2}+\lambda M^{2})}{(Q^{2}+M^{2})^2}.
\end{eqnarray}
Here $M$ is the effective mass and $\mu^{2}$ is a scale factor.
The additional parameters with their statistical errors are given
in Table II.\\
\begin{table}[h]
\caption{ The effective Parameters at low $x$ for
$0.15~\mathrm{GeV}^{2}<Q^{2}<3000~\mathrm{GeV}^{2}$ provided by
the following values. The fixed  parameters are defined by the
Block-Halzen fit to the real photon-proton cross section as
$M^{2}=0.753 \pm 0.068~ \mathrm{GeV}^{2}$, $\mu^2 = 2.82 \pm
0.290~ \mathrm{GeV}^{2}$ and $c_{0} = 0.255 \pm 0.016$ [19].}
\begin{tabular} {cccc}
\toprule \\  \multicolumn{2}{c}{parameters \quad \quad \quad ~~~~~~~~~~~~~~~~value}    \\ &&&\\ \hline \\ &&&\\
  $a_{0} $  &   \quad  $8.205\times 10^{-4}~~  \pm  4.62\times10^{-4} $  \\

  $a_{1} $  &   \quad   $-5.148\times 10^{-2}\pm 8.19\times10^{-3}$  \\

  $a_{2}$   &    \quad  $-4.725\times 10^{-3}\pm 1.01\times10^{-3}$   \\  &&&\\

 $b_{0}$   &   \quad   $2.217\times 10^{-3}\pm 1.42\times10^{-4} $ \\

 $b_{1}$   &   \quad   $1.244\times 10^{-2}\pm 8.56\times10^{-4}$  \\

 $b_{2}$    &    \quad  $5.958\times 10^{-4}\pm 2.32\times10^{-4} $ \\ &&& \\

$C_{1}$& \quad  $1.475\times 10^{-1}~\pm 3.025\times10^{-2}$ & &\\

$n$& \quad  $11.49\pm 0.99$ & &\\

$\lambda$& \quad  $2.430~\pm 0.153$ & &\\

$\chi^{2}(\mathrm{goodness~ of~ fit})$ &  \quad  $0.95$ & &\\
\hline

\end{tabular}
\end{table}
\begin{table}[h]
\centering \caption{The QCD coupling and corresponding $\Lambda$
parameter for $n_{f}=4$ at LO, NLO [19, 43] and NNLO analysis
[44].}\label{table:table2}
\begin{minipage}{\linewidth}
\renewcommand{\thefootnote}{\thempfootnote}
\centering
\begin{tabular}{|l|c|c|} \hline\noalign{\smallskip}  & $ \alpha_{s}(M_{Z}^{2})$ &$
\Lambda_{QCD}(MeV)$  \\
\hline\noalign{\smallskip}
LO & 0.1166 & 136.8 \\
NLO & 0.1166 & 284 \\
NNLO & 0.1155 & 235 \\
\hline\noalign{\smallskip}
\end{tabular}
\end{minipage}
\end{table}
\newpage{
\section{References}

1. G.G.Callan and D.Gross, Phys.Lett.B\textbf{22}, 156(1969).\\
2. L.A.Anchordoqui, C.Garca Canal and J.F.Soriano,
arXiv:1902.10134[hep-ph].\\
3. M. Klein, Ann. Phys.{\bf528}, 138 (2016).\\
4. P.Monaghan et al., Phys.Rev.Lett.{\bf110}, 152002(2013).\\
5. V.Andreev et al. [H1 Collaboration], Eur.Phys.J.C{\bf74}, 2814
(2014).\\
6. D.I.Kazakov, et.al., Phys.Rev.Lett. \textbf{65}, 1535(1990).\\
7. S.Moch, J.A.M.Vermaseren, A.Vogt, Phys.Lett.B \textbf{606},
123(2005).\\
8. M.Gluk, C.Pisano and  E.Reya, Phys.Rev.D{\bf77}, 074002(2008).\\
9. A.V.Kotikov and G.Parente, JHEP \textbf{85}, 17(1997).\\
10. A.V.Kotikov and G.Parente, Mod.Phys.Lett.A\textbf{12}, 963(1997).\\
11. G.R.Boroun and B.Rezaei, Eur.Phys.J.{\bf C72}, 2221(2012).\\
12. B.Rezaei and G.R.Boroun, Nucl.Phys.{\bf A857}, 42(2011).\\
13. G.R.Boroun, B.Rezaei and J.K.Sarma, Int.J.Mod.Phys.{\bf A29}, 1450189(2014).\\
14. N.Baruah, N.M. Nath, J.K. Sarma , Int.J.Theor.Phys.{\bf 52}, 2464(2013).\\
15. G.R.Boroun, Phys.Rev.C{\bf97}, 015206 (2018);
 B.Rezaei and G.R.boroun, Phys.Rev.C{\bf101}, 045202(2020).\\
16. L.P.Kaptari et al., JETP Lett.{\bf 109}, 281(2019).\\
17. L.P.Kaptari et al., Phys.Rev.{\bf D99}, 096019(2019).\\
18. M.M.Block et al., Phys.Rev.D{\bf84}, 094010 (2011).\\
19. M.M.Block et al., Phys.Rev.D{\bf89}, 094027 (2014).\\
20. G.Altarelli and G.Martinelli, Phys.Lett.B\textbf{76}, 89(1978).\\
21.Yu.L.Dokshitzer, Sov.Phys.JETP {\textbf{46}}, 641(1977);
G.Altarelli and G.Parisi, Nucl.Phys.B \textbf{126}, 298(1977);
V.N.Gribov and L.N.Lipatov, Sov.J.Nucl.Phys. \textbf{15},
438(1972).\\
22. A.Donnachie and P.V.Landshoff, Phys.Lett.B {\bf437}, 408(1998 ).\\
23. A.Donnachie and P.V.Landshoff, Phys.Lett.B {\bf550}, 160(2002 ).\\
24. B.G. Shaikhatdenov, A.V. Kotikov, V.G. Krivokhizhin, G.
Parente, Phys. Rev. D {\bf81}, 034008(2010).\\
25. M.Devee, R.Baishya and J.K.sarma, Eur.Phys.J.C{\bf72},
2036(2012); M.M.Block et al., Phys.Rev.D{\bf83}, 054009(2011); M.M.Block et al., Eur.Phys.J.C{\bf69}, 425 (2010);
A.Cafarella et al., Nucl.Phys.B{\bf748}, 253(2006);  A.Cafarella et al., Comput.Phys.Commun.{\bf179}, 665(2008).\\
26. G.R.Boroun and B.Rezaei, Eur.Phys.J.C{\bf73}, 2412(2013);
 S.Zarrin and G.R.Boroun, Nucl.Phys.{\bf B922}, 126 (2017);
 F. Taghavi-Shahri, A. Mirjalili and M.M. Yazdanpanah , Eur.Phys.J.{\bf C71}, 1590(2011); H.Khanpour et al., Phys.Rev.C{\bf95}, 035201(2017).\\
27. E. Floratos, C. Kounnas and R. Lacaze, Nucl. Phys. B
\textbf{192}, 417(1981);  J. Blumlein, V. Ravindran and W. van
Neerven, Nucl. Phys. B \textbf{586}, 349(2000).\\
28. C.G. Callan, Jr., Phys. Rev. D{\bf2}, 1541 (1970); K.
Symanzik, Comm. Math. Phys.{\bf18}, 227 (1970), Comm. Math. Phys.
B{\bf39},49 (1971);
G. Parisi, Phys. Lett. B{\bf39}, 643 (1972).\\
29. W.L. van Neerven, A.Vogt, Phys.Lett.B \textbf{490}, 111(2000).\\
30. A.Vogt, S.Moch, J.A.M.Vermaseren, Nucl.Phys.B \textbf{691}, 129(2004).\\
31. A.A.Godizov,  Phys.Rev.{\bf D96}, 034023 (2017).\\
32. M.M.Block et al., Phys.Rev.D{\bf77}, 094003 (2008).\\
33. A.D.Martin, W.J.stirling and R.S.Thorne, Phys.Lett.B{\bf635}, 305(2006).\\
34. K Golec-Biernat and A.M.Stasto, Phys.Rev.D {\bf80},
014006(2009).\\
35. B.Rezaei and G.R.boroun, Eur.Phys.J.A{\bf55}, 66(2019); G.R.Boroun, Eur.Phys.J.Plus
{\bf135}, 68(2020).\\
36. N.N.Nikolaev and B.G.Zakharov, Phys.Lett.{\bf B332}, 184(1994);   N.N.Nikolaev and V.R.Zoller,  Phys.Atom.Nucl.{\bf 73}, 672 (2010).\\
37. B.Rezaei and G.R.boroun, Phys.Lett.{\bf B692}, 247(2010).\\
38. M.Froissart, Phys.Rev.{\bf123}, 1053 (1961).\\
39. M.M.Block et al., Phys.Rev.D{\bf88}, 014006 (2013).\\
40. M. M. Block and F. Halzen, Phys. Rev. D{\bf70},
091901(2004).\\
41. F. D. Aaron et al. [H1 and ZEUS Collaborations], JHEP{\bf
1001}, 109 (2010).\\
42. A. Y. Illarionov, A. V. Kotikov and G. Parente Bermudez, Phys.
Part. Nucl. {\bf39}, 307 (2008).\\
43. S. Chekanov et al. [ZEUS Collaboration], Eur. Phys. J. {\bf C21}, 443 (2001).\\
44. A.D.Martin et al., Phys.Letts.{\bf B604}, 61(2004).\\
45. M.Niedziela and M.Praszalowicz, Acta Phys.Polon. B{\bf46}, 2019(2015).\\
46. C.Ewerz et al., Phys.lett.B\textbf{720}, 181(2013).\\
47.  V.Tvaskis et al., Phys.Rev.{\bf C97}, 045204(2018);
Phys.Rev.Lett. 98 (2007) 142301.\\
48. F.D. Aaron et al. [H1 Collaboration], phys.Lett.B\textbf{665},
139(2008); Eur.Phys.J.C\textbf{71},1579(2011).\\
49. C.Ewerz and O.Nachtmann, Phys.Lett.B\textbf{648}, 279(2007).\\
50. H.Abromowicz et al. [ZEUS Collaboration],
Phys.Rev.D\textbf{9}, 072002(2014).\\}
\begin{figure}
\includegraphics[width=0.55\textwidth]{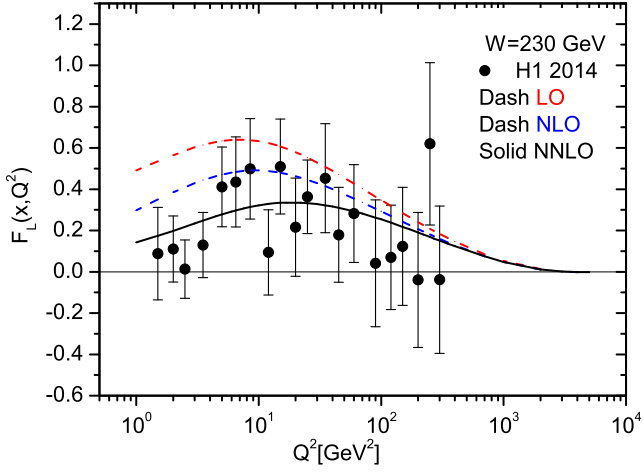}
\caption{The longitudinal structure function extracted at LO upto
NNLO in comparison with the H1 data [5] as accompanied with total
errors. The results are presented  at fixed value of the invariant
mass W  in the interval $1~\mathrm{GeV}^{2 }\leq Q^{2} \leq
3000~\mathrm{GeV}^{2}$ at low values of $x$.}\label{Fig1}
\end{figure}
\begin{figure}
\includegraphics[width=0.78\textwidth]{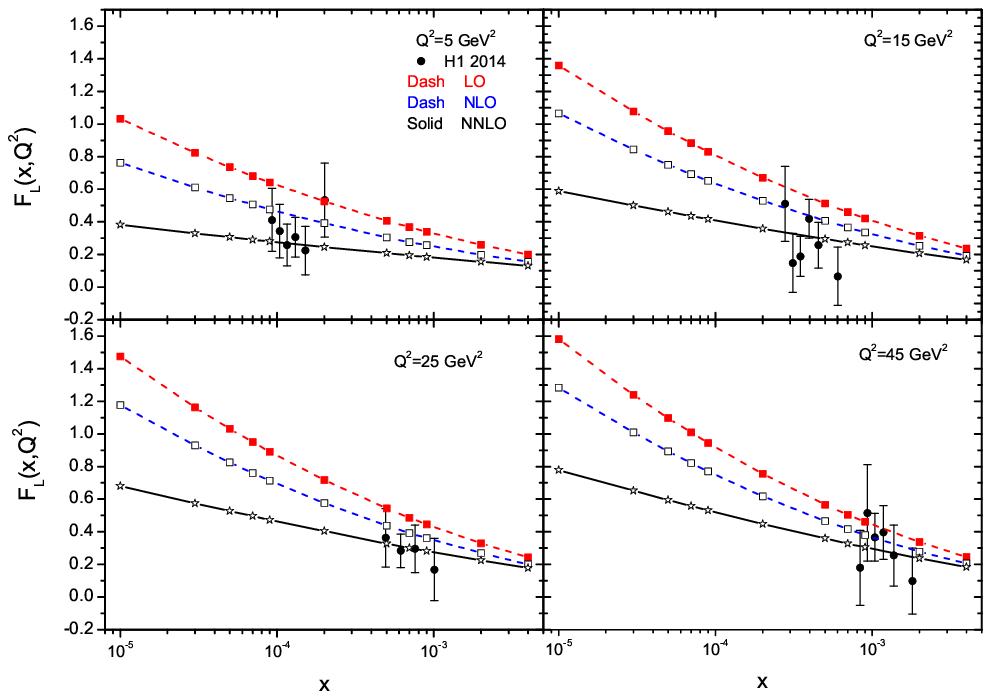}
\caption{Predictions of $F_{L}(x,Q^{2})$ at four $Q^{2}$  values
$5,15,25$ and $45~\mathrm{GeV^{2}}$ at LO upto NNLO, compared with
the H1 data [5].  Dash lines are the longitudinal structure
function at LO (full square ) and  NLO (open square). Solid line
is the longitudinal structure function at NNLO approximation(open
star). }\label{Fig2}
\end{figure}
\begin{figure}
\includegraphics[width=0.55\textwidth]{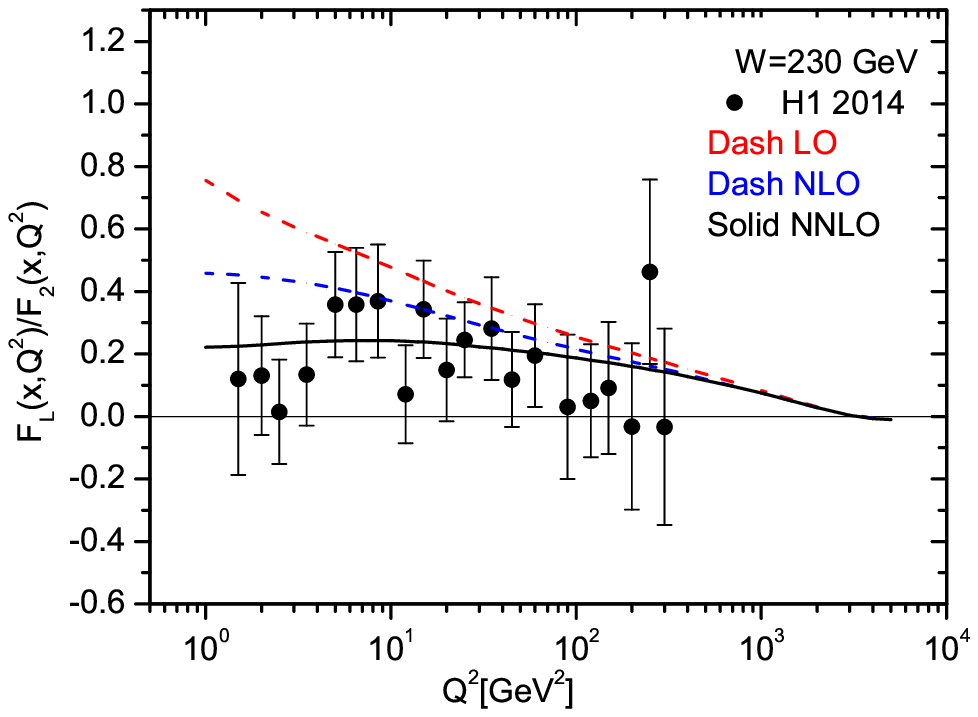}
\caption{Ratio $F_{L}/F_{2}$ plotted as function of $Q^{2}$
variable at $W=230~\mathrm{GeV}$ compared with the H1 data [5].
}\label{Fig3}
\end{figure}
\begin{figure}
\includegraphics[width=0.8\textwidth]{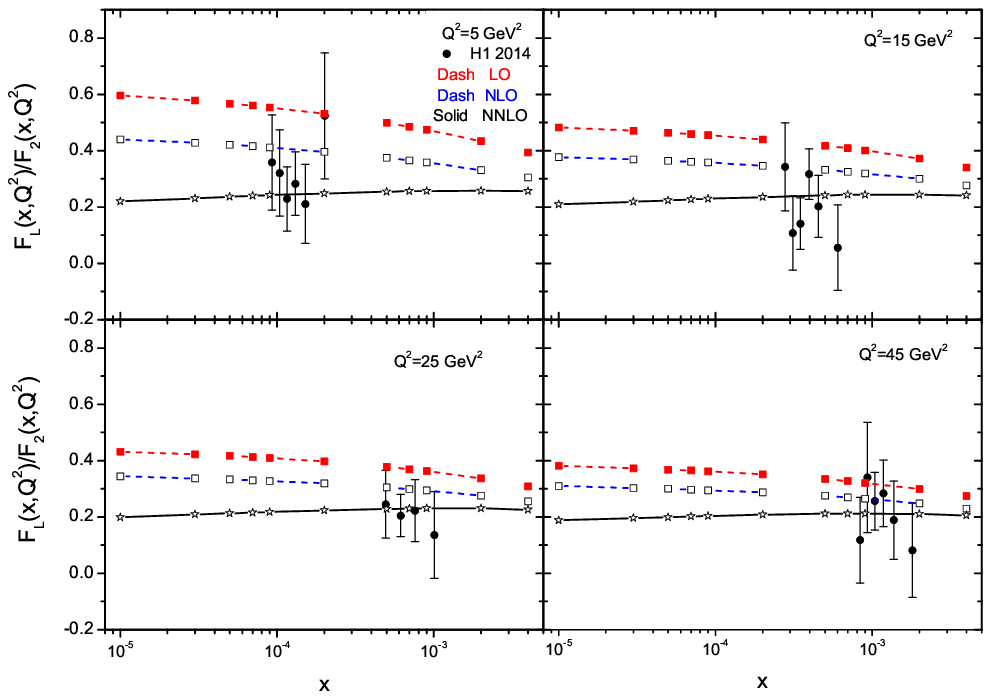}
\caption{Predictions of the ratio $F_{L}/F_{2}$ at four $Q^{2}$
values $5,15,25$ and $45~\mathrm{GeV^{2}}$ at LO upto NNLO,
compared with the H1 data [5]. Dash lines are the longitudinal
structure function at LO (full square ) and  NLO (open square).
Solid line is the longitudinal structure function at NNLO
approximation(open star).}\label{Fig4}
\end{figure}
\begin{figure}
\includegraphics[width=0.55\textwidth]{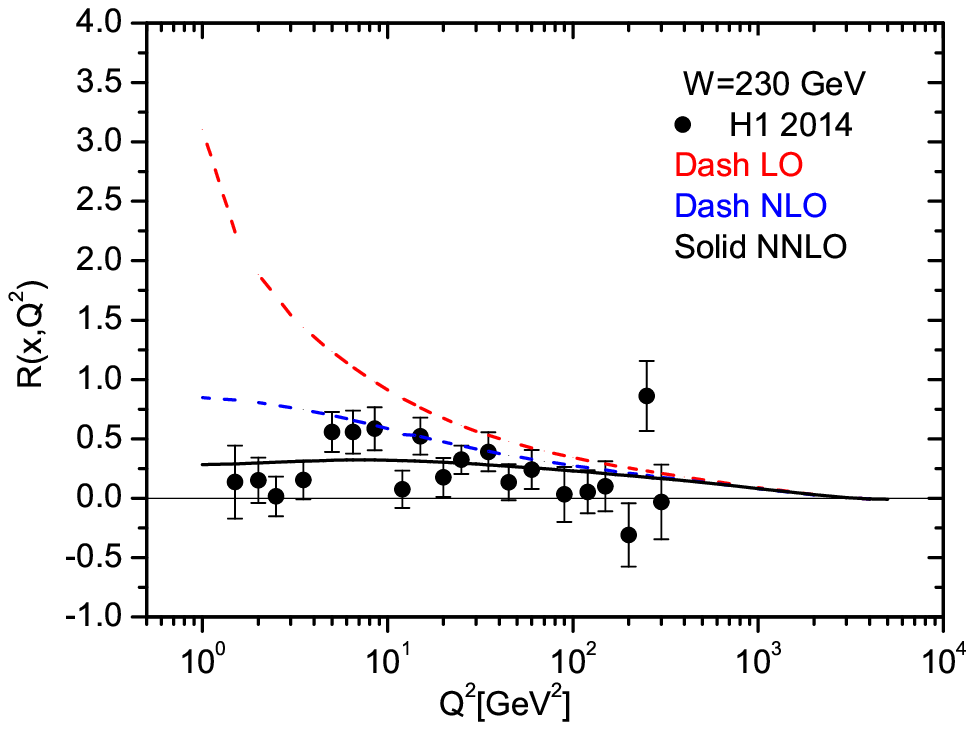}
\caption{Ratio $R(x,Q^{2})$ plotted as function of $Q^{2}$
variable at $W=230~\mathrm{GeV}$ compared with the H1 data [5].
}\label{Fig5}
\end{figure}
\begin{figure}
\includegraphics[width=0.8\textwidth]{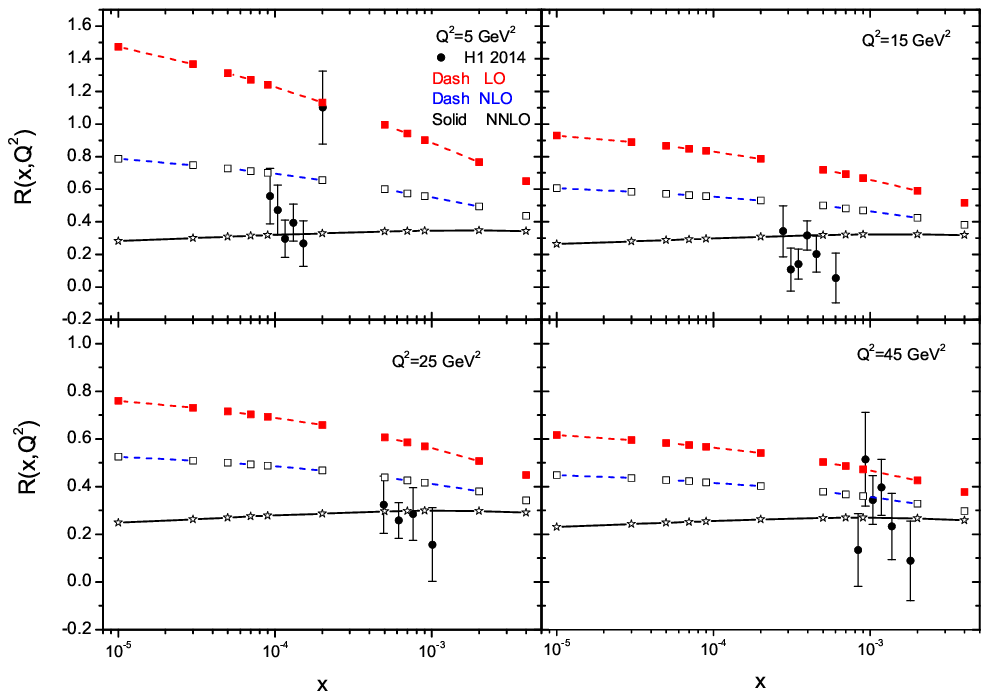}
\caption{Predictions of the ratio $R(x,Q^{2})$ at four $Q^{2}$
values $5,15,25$ and $45~\mathrm{GeV^{2}}$ at LO upto NNLO,
compared with the H1 data [5]. Dash lines are the longitudinal
structure function at LO (full square ) and NLO (open square).
Solid line is the longitudinal structure function at NNLO
approximation (open star).}\label{Fig6}
\end{figure}
\begin{figure}
\includegraphics[width=0.55\textwidth]{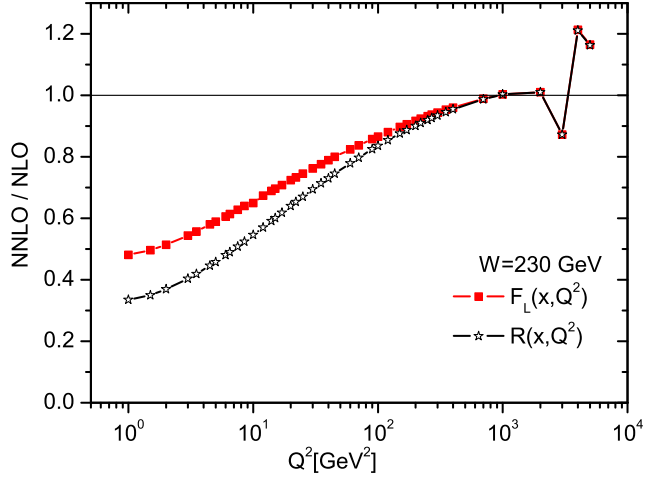}
\caption{The NNLO correction factor for $F_{L}(x,Q^{2})$ (full
square) and $R(x,Q^{2})$ (open star) in the interval
$1~\mathrm{GeV^{2}}<Q^{2}<5000~\mathrm{GeV^{2}}$ at fixed value of
the invariant mass $W=230~\mathrm{GeV}$.}\label{Fig7}
\end{figure}
\begin{figure}
\includegraphics[width=0.8\textwidth]{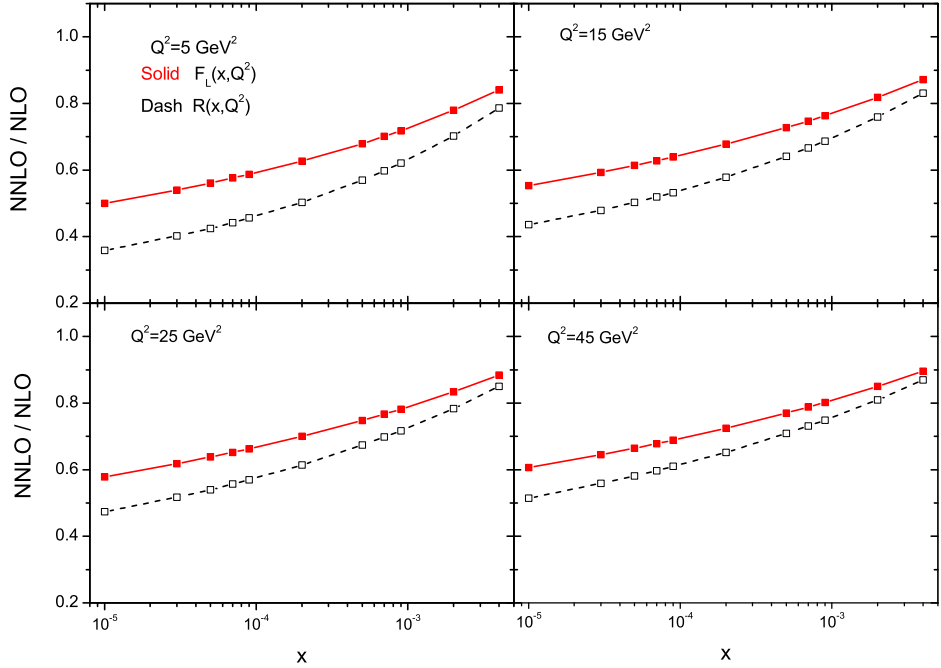}
\caption{Ratio of the NNLO to NLO contributions to
$F_{L}(x,Q^{2})$ (full square) and $R(x,Q^{2})$ (open square) for
$Q^{2}=5,15,25$ and $45~\mathrm{GeV^{2}}$. }\label{Fig8}
\end{figure}

\end{document}